\documentclass[a4paper,11pt]{article}
\pdfoutput=1 

\usepackage{jcappub} 

\usepackage[T1]{fontenc} 
\newcommand{\hex}{$(^4$He$\Xmm)$}
\def\Xmm{X^{--}}

\title{\boldmath BBN catalysis by doubly charged particles}

\author[a]{Evgeny Akhmedov}
\author[b,c]{and Maxim Pospelov}

\affiliation[a]{Max-Planck-Institut f\"{u}r Kernphysik, Saupfercheckweg 1,
69117 Heidelberg, Germany}
\affiliation[b]{William I. Fine Theoretical Physics Institute,  
School of Physics and Astronomy,\\ University of Minnesota, 
Minneapolis, MN 55455, USA}
\affiliation[c]{
School of Physics and Astronomy, University of
Minnesota,\\ Minneapolis, MN 55455, USA} 
\emailAdd{akhmedov@mpi-hd.mpg.de}
\emailAdd{pospelov@umn.edu}

\abstract{
We consider primordial nucleosynthesis in the presence of hypothetical
quasi-stable doubly charged particles. Existence of $\Xmm$ with
macroscopic lifetimes will lead to the formation of its bound states with
$^4$He and other
light elements, significantly facilitating the subsequent
formation of lithium nuclei. From observational constraints on maximum
allowable amount of lithium, that we update in this work, we derive
strong constraints on the abundance and lifetime of $\Xmm$. In a
likely cosmological freeze-out scenario with temperatures initially
exceeding the mass of $\Xmm$, the BBN constrains the lifetime of these
particles to be less than about 100 seconds. For parametrically long
lifetimes, lithium abundance data constrain $\Xmm$ abundance to be less
than $10^{-9}$
relative to protons, regardless of whether these particles
decay or remain stable. Stable particles could saturate the dark matter
density only if their mass is comparable to or in excess of
$10^{10}$\,GeV, and most of $\Xmm$ will be found in bound states with
beryllium nuclei, so that chemically they would appear as abnormally
heavy helium isotopes.
}

\begin{document}
\maketitle
\flushbottom

\section{Introduction} 

Metastable charged particles occur in many extensions of the Standard 
Model (SM) of particle physics.
If the charge of such beyond-SM (BSM) particles is not small compared to 
the electron charge,
they are guaranteed to be at or above the electroweak scale, given 
the capacity of modern experiments at the LHC as well as past experiments at
LEP and the Tevatron. What sets charged particles in a completely different 
category from any BSM neutral metastable particle is the plethora 
of new phenomena that the charge implies. These additional signatures 
afford new experimental pathways for searches for such particles, see 
{\em e.g.} the review \cite{Perl:2001xi}.

Chiefly among important experimental signatures for charged metastable 
particles $X$ 
stands the presence of ionization tracks. For particles that 
interact solely due to electromagnetic interactions, the slow-down 
process is very inefficient, $|dE_X/dx|\sim |dE_\mu/dx|$, and given a 
large reservoir of kinetic energy, this results in long persistent 
ionizing tracks. The search for such tracks is performed both at ATLAS 
and the CMS experiments \cite{ATLAS:2019gqq,CMS:2016kce}, resulting in 
sensitivity to $X$ at the level of a few hundred GeV (assuming standard 
electromagnetic or electroweak pair production mechanisms). The presence 
of charge also gives reasons to believe that ``almost" stable particles can be 
accelerated at astrophysical sites and can be present as a rare component of 
cosmic rays (CR). This opens up the possibility to extend such searches much 
beyond the electroweak scale \cite{Perl:2001xi}, modulo of course the 
assumption about the overall abundance of such particles.

Negatively charged massive particles $X^-$ are of special interest. It 
is well known that they can form bound states with nuclei, thereby 
reducing the Coulomb barrier and facilitating nuclear reactions. 
Muon-catalyzed fusion reactions have been investigated over a number of 
years \cite{Breunlich:1989vg}, and some exotic BSM possibilities for the  
catalysis were also discussed \cite{Hamaguchi:2012px}. The two limiting 
factors that complicate any practical use of muon catalysis are the 
relatively short muon lifetime and formation of bound states 
with $Z\geq2$ nuclei that ``hide" $\mu^-$ from subsequent participation 
in nuclear fusion. Recently, it was argued that a heavy BSM {\em 
doubly-charged} particle $X^{--}$ would offer a huge advantage over $X^-$ and 
{would potentially be able to catalyze enough reactions to achieve 
positive energy balance, {\em i.e.} have energy release in excess of the 
energy needed for production or acquisition of such a particle  
\cite{Akhmedov:2021qmr}. 
Despite a rather futuristic nature of such investigations, it is important to 
have an example of a ``useful" BSM particle, as this may give a new 
impetus to collider and CR searches of $X^{--}$. Incidentally, a recent 
ATLAS analysis \cite{ATLAS:2022pib,ATLAS:2024fdw} showed some hints of  
the ionization tracks at the LHC with $|dE/dx|$ larger than that   
for muons, prompting the speculations that such tracks correspond to exotic 
$X^{--}$ or $X^{++}$ \cite{Giudice:2022bpq,Akhmedov:2024rvp}.

One of the most important defining properties of a charged BSM particle is 
its lifetime. The LHC studies could potentially determine that the 
lifetime of $X$ along the anomalous track is in excess of {\em e.g.} 
10 ns. Cosmic ray searches on the other hand would be sensitive to 
particles with lifetimes $>{\cal O}({\rm years})$. This leaves many 
orders of magnitude in between, where only indirect studies are 
possible. The purpose of this work is to investigate the constraints on 
the lifetime and abundance of $X^{--}$ imposed by cosmology, and 
specifically by the Big Bang Nucleosynthesis (BBN). It is well known 
that the presence of singly charged $X^{-}$ with lifetimes in excess 
of ${\cal O}(10^3\,{\rm s})$ in the early Universe leads to the catalysis of 
certain nuclear reactions and increased yield of such rare elements as $^6$Li 
(the scenario known as CBBN) \cite{Pospelov:2006sc}. Observations of these 
elemental abundances, in turn, set strong constraints on the early 
Universe abundance of $X^{-}$, {\em irrespective} of subsequent fate of 
$X^{-}$ (whether it decays or remains stable). Of course, if a particle 
decays, more constraints on energy injection during the BBN and cosmic 
microwave background (CMB) epochs would 
apply (see {\em e.g.} BBN review in \cite{Workman:2022ynf} and references 
therein). The energy injection constraints are, however, quite model 
dependent. The BBN catalysis by $X^{--}$ and consequent 
constraints on the abundance and lifetime of such a particle have not been 
studied thus far, and our goal is to fill 
this gap. We note that in addition to the motivations stated above, 
stable $X^{--}$ particles that form bound states with helium and beryllium  
nuclei have been suggested as dark matter (DM) candidates in a number of 
publications \cite{Fargion:2005ep,Cudell:2014dva}. Our analysis on the 
abundance of $X^{--}$ during the BBN would then provide a 
{\em lower limit} for the mass of such DM particles.

If the lifetimes of the doubly charged particles are comparable to or 
longer than the age of the Universe, such remnants could still be 
present, in particular, in the solar system, and on Earth. A doubly 
positively charged particle $X^{++}$ would appear as an abnormally heavy 
helium, and in a certain mass range its abundance will be limited by the 
negative results of searches for anomalously heavy helium isotopes 
\cite{Mueller:2003ji}. 
In addition, the BBN-generated yield of $({\rm Be}\Xmm)$ bound states%
\footnote{Throughout this paper we use the brackets to denote states bound 
by the Coulomb force.} will also be seen as abnormal helium, and will be 
present even if the evolution of the “$X$-sector” is charge-asymmetric,
so that $X^{++}$ are not currently around. Moreover, the bound state 
of $\Xmm$ with lithium isotopes will be chemically equivalent to 
hydrogen, and in a certain mass range will be subject to the 
very sensitive searches for 
anomalously heavy hydrogen performed long time ago 
\cite{Smith:1982qu,Verkerk:1991jf}. Determining the yields of (Be$\Xmm$) and 
(Li$\Xmm$) is another goal of our paper.

The existence of doubly-charged particles in BSM models is not 
guaranteed. Indeed, the SM itself does not contain elementary doubly 
charged states (as opposed to composite $\Delta^{++}$ baryons and 
negatively charged antibaryons $\bar \Delta^{--}$). While of course 
$X^{\pm\pm}$ can be added to any theory ``by hand", there is a 
number of BSM models where doubly charged particles make natural 
appearance. 
These include Georgi-Mahacek model, Zee-Babu model, little Higgs model 
and others, as surveyed in Ref. \cite{Akhmedov:2021qmr}. How could  
metastability of charged BSM particles and their sizeable abundance be 
achieved? A recent example was presented in \cite{Akhmedov:2024rvp}, 
based on the left-right symmetric models \cite{Mohapatra:1979ia}, and 
the triplet set of Higgses as an important ingredient. Among such 
triplet states, there are {\em elementary} di-charged Higgs states 
$\Delta^{\pm\pm}_R$. As shown in \cite{Akhmedov:2024rvp}, a combination 
of the discrete symmetry $\Delta_R \to - \Delta_R$ and a certain 
hierarchy between the masses of BSM particles, $m_{\Delta_R} \ll m_{W_R}$, 
will lead to metastability of the doubly charged Higgs. Supersymmetric 
models, where there is a plethora of BSM charged particles, offer a general 
pathway to metastability 
of charged particles. The following mass 
arrangement of superpartners gives the requisite longevity within the 
context of a supersymmetric SM: gravitino as the lightest (LSP), and 
{\em e.g.} a charged superpartner of a SM lepton as the next-to-lightest 
superpartner (NLSP). With such mass arrangement, NLSP$\to$LSP decay is 
suppressed by the small coupling, with the decay rate suppressed by the 
Planck mass. This way, the lifetime of the 
charged NLSP can vary from seconds to many years, giving enough time for 
the BBN catalysis to occur.  If, in addition, a supersymmetric model 
contains doubly charged states, then the superpartners of $X^{\pm\pm}$ 
({\em e.g.} a doubly charged Higgsino $\tilde 
\Delta^{\pm\pm}_R$) can indeed be long-lived if it is the NLSP, and 
gravitino is the LSP. This longevity mechanism relies, of course, on the 
existence of the discrete $R$-parity. As to the significant abundance of 
the doubly charged states, this is relatively easy to arrange. For 
example, if the reheating scale is sufficiently large, reaching 
$T_{rh}\sim m_X$,\footnote{We use the units $\hbar=c=k_B=1$ throughout this 
paper.}
 these particles are guaranteed to be thermalized, with 
subsequent freeze-out via $X^+X^-\to$\,SM that will provide sufficient 
abundances to be probed via CBBN.

This paper is organized as follows. In the next section we discuss 
observational constraints on Li/Be/B abundances and quote the values 
that we will take as upper limits on primordial abundances. In section 
3 we determine the properties of bound states of $X^{--}$ with light 
nuclei and identify the important nuclear reactions. In section 4 we 
perform calculations, and also recast some known results,  
to determine the rates for the most important catalyzed reactions. In section 
5 we perform the CBBN analysis of the Li/Be/B yields and set constraints on 
the abundance of $X^{--}$ in a wide range of its possible lifetime. 
Section 6 derives abundances of anomalously heavy hydrogen and helium 
({\em i.e.} of the corresponding bound states of nuclei with $\Xmm$).  
Section 7 contains the discussion and our conclusions.

\section{Observational constraints of primordial Li/Be/B} 

Abundances of lithium, beryllium and boron are typically very small, 
compared not only with the main primordial elements such as hydrogen and 
helium, but also 
with star-produced elements, such as carbon and 
oxygen. Therefore, if the post-BBN evolution of these elements is 
relatively mundane, one can use their observations as a strong 
constraint on scenarios that deviate from the minimal (or standard) BBN 
(SBBN). This section contains a discussion  
of the values of primordial abundances of 
these elements that will be adopted in our study. 
We will specifically concentrate on these rare elements, as the catalytic 
influence of heavy charged relics on deuterium and helium is relatively 
mild \cite{Pospelov:2006sc}.

With the rapid progress of precision cosmology in the last two decades, 
we now have a very accurate determination of the baryon-to-photon ratio 
$\eta_B$, that for a long time was one of the main uncertainty factors 
influencing the BBN predictions. Observations of D/H, as well as of 
primordial helium, are broadly consistent with the BBN predictions, at a 
few percent level. The comparison of observations of lithium isotopes 
with the corresponding SBBN predictions tells a different, and perhaps 
rather confusing, story.

Stable isotopes of lithium, $^6$Li and $^7$Li, are produced in rather 
different amounts in SBBN. At current values of $\eta_B$, $^7$Li is 
produced predominantly as $^7$Be, that much later decays to lithium via 
the electron capture reaction. $^7$Be is produced via the mostly 
$E1$-dominated $(\alpha,\gamma)$ reaction on $^3$He. Once produced, 
$^7$Be proves to be rather robust against burning by protons, 
and its reduction by neutrons is also minimal. In contrast, the yield of 
$^6$Li is several orders of magnitude smaller, 
for two reasons. The $(\alpha,\gamma)$ reaction leading to helium-deuterium 
fusion is very suppressed in the $E1$ channel on account of nearly same 
charge-to-mass ratio for D and $^4$He nuclei. In addition, $^6$Li 
undergoes very efficient burning by protons, that remains active down to 
very small temperatures, $T\sim 10$\,keV. Thus, the SBBN determines the 
resulting yields to be \cite{Yeh:2020mgl}
\begin{eqnarray}
\label{Li7SBBN}
   \left. {\rm \frac{^7Li}{H} }\right|_{\rm SBBN} &=&  (4.94 \pm 0.72) \times 
    10^{-10},\\
   \left. {\rm \frac{^6Li}{H} }\right|_{\rm SBBN} &\simeq& 10^{-14} .
   \label{Li6SBBN}
\end{eqnarray}
Other evaluations are in agreement with (\ref{Li7SBBN}) within errors 
\cite{Consiglio:2017pot,Pitrou:2020etk}.

Comparison with observations is, however, far from being 
straightforward. First of all, some amount of lithium is produced by 
CR, and the expected amount of this production mechanism, 
albeit with large uncertainties, by far exceeds prediction 
(\ref{Li6SBBN}). The presence of CR predating main chemical 
evolution of elements in late Universe will produce $^6$Li uncorrelated 
with metallicity, and therefore prediction (\ref{Li6SBBN}) has no 
relevance for comparison with observations. It has to be said that the 
initial excitement about the claims of metallicity-uncorrelated and 
possibly nonstandard-BBN-induced $^6$Li \cite{Asplund:2005yt} turned out 
to be short-lived, as these claims have been negated by subsequent 
re-analyses.

Observations of \,$^7$Li are telling another, and quite a different story. 
The relative constancy of $^7$Li/H in the atmospheres of old and hot 
metal-poor stars in the Milky Way was discovered in the 1970s, and for a 
long time believed to be a true representation of the primordial lithium 
abundance \cite{Walker:1991ap}. This might be the case if during the 
$\sim$13 billion years of these stars' existence there has been no 
significant depletion of primordial lithium in their atmospheres. 
Interpreted this way, the so-called Spite plateau abundance 
\cite{Spite:1982dd} is in sharp contradiction with the SBBN prediction 
(\ref{Li7SBBN}), exhibiting a factor of 3 deficit. Latest values, as 
quoted in \cite{Fields:2022mpw} and references therein, are given by
\begin{eqnarray}
   \left. {\rm \frac{^7Li}{H} }\right|_{\rm Spite~plateau} = (1.6\pm 0.3)
   \times 10^{-10}.
   \label{Spite}
\end{eqnarray}
This discrepancy is often referred to as ``primordial lithium problem", 
and it has generated a great scrutiny of all its aspects, 
starting from re-evaluation of nuclear reaction rates to critical 
assessment of the assumption about lithium preservation in the 
atmospheres of old stars, or even various beyond-SBBN scenarios (see 
{\em e.g.} discussions in 
\cite{Cyburt:2008kw,Jedamzik:2009uy,Pospelov:2010hj,Fields:2011zzb,
Fields:2022mpw}).

While this is difficult to prove outright, there is a growing consensus 
that indeed the diffusion and gravitational settling of lithium in the 
atmospheres of old Population II stars may be responsible for a factor 
of $\sim 3$ depletion of lithium, from (\ref{Li7SBBN}) to (\ref{Spite}). 
Recently, it has been argued \cite{Fields:2022mpw} that non-observation 
of $^6$Li in amounts expected to be generated by the CR activity lends 
some support towards stellar depletion of both isotopes of lithium. 
While this may solve the ongoing discrepancy between (\ref{Li7SBBN}) and 
(\ref{Spite}), it does not answer the question of what primordial 
amounts of lithium current data allow. In this paper, while not 
attempting to address this important question in its full entirety, we 
try to determine the maximum amounts of $^6$Li/H and of $^7$Li/H 
that can be still consistent with observations. We will then contrast them  
with our CBBN predictions to derive robust limits on our model. 
Our interest in mass-6 isotope stems from the fact that in the CBBN version 
we investigate  in this paper it is the catalysis of $^6$Li that is 
most pronounced, in the limit of long lifetime for $\Xmm$. 
In carrying out the analysis, one may assume that the abundance of 
$^7$Li is set by the standard processes -- SBBN production, CR and 
stellar production at high metallicities, as well as possible stellar 
depletion. Within these assumptions, we want to {\em maximize} 
$^6$Li/H$|_{\rm BBN}$ that can then be treated as a reasonable upper bound 
for the catalyzed BBN scenario.

First, we note that it is not easy to derive an upper bound on 
$^6$Li/H$|_{\rm BBN}$ using the constraints on the abundance of 
$^6$Li in the atmospheres of old stars. If $^7$Li/H was indeed depleted by a 
a factor of ${\cal O}(3)$, then $^6$Li/H could have been depleted by a much 
greater factor, due to its 
fragility. We do not know of any reasonable way of estimating such 
a factor without detailed modelling of lithium diffusion/settling in 
the atmospheres of metal-poor stars.

An alternative route for inferring the primordial abundances that avoids 
Spite-type determination altogether and uses the interstellar medium (ISM)
abundances of lithium and $^6$Li/$^7$Li ratio has been touted (see 
{\em e.g.} Refs.~\cite{Kawanomoto:2003tlp,Knauth:2002dd,Prodanovic:2004jw,Kawanomoto}) as 
a difficult but completely independent from stellar uncertainties method 
for inferring primordial lithium abundances. While most of 
the observations were performed as studies of absorption along the line 
of sight towards the direction of bright nearby stars \cite{Knauth:2002dd,
Kawanomoto}, 
observations of interstellar gas 
outside the Milky Way (in the Small Magellanic Cloud, or 
SMC), at quarter of the solar 
metallicity, have also yielded measurements of 
$^7$Li/H and $^6$Li/$^7$Li \cite{Howk:2012rb}.

We shall consider two approaches to constraining $^6$Li/H$|_{\rm BBN}$. 
The first one refers to the absolute abundances of $^6$Li/H$|_{\rm obs}$ 
measured in the solar system \cite{Anders:1989zg}, and also (with larger 
uncertainty) in the 
ISM \cite{Howk:2012rb}. Using these values, we can build a series of 
inequalities,
\begin{eqnarray}
\left. {\rm \frac{^6Li}{H} }\right|_{\rm BBN} <  \left. {\rm \frac{^6Li}{H} }
\right|_{\rm BBN} + \left. {\rm \frac{^6Li}{H} }\right|_{\rm CR} < 
{\cal F}_{\rm max} \left. \times \frac{^6\rm Li}{\rm H} \right|_{\rm obs}.
\end{eqnarray}
Here ${\cal F}$ stands for a fudge suppression factor, ${\cal F}>1$, 
that relates the initial, BBN plus CR, abundance of $^6$Li and the observed 
abundance: 
\begin{eqnarray}
   \left.  \frac{^6\rm Li}{\rm H} \right|_{\rm obs} = 
   \frac{1}{{\cal F}}\times \left( \left. {\rm \frac{^6Li}{H} }
\right|_{\rm BBN} + \left. {\rm \frac{^6Li}{H} }\right|_{\rm CR}\right). 
\end{eqnarray}
The factor ${\cal F}$ can be written as ${\cal F = AD}$, where 
${\cal A}$ accounts for the suppression due to astration 
(cycling of the primordial gas in stars), which leads to 
the physical destruction of $^6$Li, and ${\cal D}$ is the factor that takes 
into account adsorption of $^6$Li on grains of dust,  
also leading to the reduced values of $^6$Li. 
To set conservative constraints on primordial $^6$Li, we need to 
{\em maximize} the total suppression factor 
${\cal F}$ by maximizing possible astration and adsorption.

For observations inside the solar system, such as in meteorites, we 
believe that the correction for the adsorption on dust is not 
particularly relevant. We can get an idea of how much stellar recycling 
of the initial gas has occurred from observations of deuterium in the 
vicinity of the solar system. Deuterium is even more fragile than 
$^6$Li, but its total suppression compared to the initial BBN value due 
to adsorption/astration is never larger than by a factor of 1.5 
\cite{Geiss1998239,Friedman}. This implies that we can safely assume 
${\cal F}_{\rm max,\odot} < 2$, resulting in the following upper 
bound:
\begin{eqnarray}
    \left. {\rm \frac{^6Li}{H} }\right|_{\rm BBN} <  2 \times \left. 
{\rm \frac{^6Li}{H} }\right|_{\rm \odot} = 3 \times 10^{-10}.
    \label{limit_Sun}
\end{eqnarray}

Aside from the observations of lithium in stars and in the solar system, 
there are a few observations of lithium in the ISM. 
The recent measurement of $^7$Li/H in SMC, ${\rm ^7Li/H}|_{\rm SMC}=4.8\times 
10^{-10}$, also makes a tentative determination of lithium isotopic 
ratio and concludes that ${\rm ^6Li/^7Li}|_{\rm SMC}<0.28$ at $3\sigma$ 
\cite{Howk:2012rb}. Comparison with other elements (K, S) in SMC and solar 
system allows to calibrate away the effect of the dust, and in that sense we 
may take into account only the astration factor, that we again assume to be 
not larger than 2 
\footnote{In this respect, observations of D/H in the same systems where 
${\rm ^6Li/^7Li}$ and ${\rm Li}/$H are measured are desirable.}. 
This way we derive
\begin{eqnarray}
 \left. {\rm \frac{^6Li}{H} }\right|_{\rm BBN} <  2 \times \left. 
     {\rm \frac{^6Li}{H} }\right|_{\rm SMC} = 2.7 \times 10^{-10},
    \label{limit_SMC}
\end{eqnarray}
which is a nearly the same as the value in (\ref{limit_Sun}).  

Similar ballpark numbers can be achieved from older measurements of both 
isotopes of lithium in the ISM in proximity to solar system. However, 
the scatter in the results is larger than that for the measurements 
discussed above. Therefore, we will adopt (\ref{limit_Sun}) as a reliable and 
conservative limit on the maximal abundance of $^6$Li that can emerge 
from the BBN that is consistent with solar system and ISM measurements.

One can also use solar system measurements, $^7$Li/H in conjunction with D/H, 
to set the constraint on the maximum possible production of $^7$Li 
in the (non-standard) BBN. This way, we get a very relaxed, and 
therefore quite conservative bound,
\begin{eqnarray}
   \left. {\rm \frac{^7Li}{H} }\right|_{\rm BBN} <  2 \times \left. 
     {\rm \frac{^7Li}{H} }\right|_{\rm \odot} = 4 \times 10^{-9}.
    \label{limit_Sun_Li7} 
\end{eqnarray}
We note that this value is about an order of magnitude 
larger than the SBBN one, as $^7$Li in the vicinity of the solar 
system must have resulted from the prior CR and stellar activity. Our 
limit (\ref{limit_Sun_Li7}) assumes that all of it in fact came from the 
BBN, making it a rather conservative assumption for the purposes of 
limiting non-standard BBN scenarios.

Beryllium and boron are also extremely rare elements. The smallest 
values are found at lowest metallicities, conforming with the 
expectations of their secondary (not BBN) origin. Observations of these 
elements do not show any sign of Spite-like plateau, and if one can neglect 
depletion of these elements, one could derive a constraint on (nonstandard) BBN 
abundances of Be and B. In deriving these approximate limits, we take 
into account the possibility of an ${\cal O}(3)$ suppression of their 
abundance in stellar atmospheres, as for $^7$Li. (Unlike $^6$Li, these 
elements are less fragile than $^7$Li). Taking that into account and using 
measurements quoted in \cite{Fields:2022mpw}, we can conservatively 
conclude that
\begin{eqnarray}
    \left. {\rm \frac{^9Be}{H} }\right|_{\rm BBN} < 5\times 10^{-13} ; 
~~ \left. {\rm \frac{B}{H} }\right|_{\rm BBN} < 10^{-10}, 
\end{eqnarray}
where B is understood as $^{10}$B+$^{11}$B. BBN contributions in excess of 
these values would generate Spite-like plateaus in the data, that is not seen. 
As we shall see, in the CBBN scenario with $X^{--}$ the yields for these 
elements may be enhanced by many orders of magnitude compared with those in 
SBBN.

\section{Properties of bound states and recombination rates} 

In this section we will investigate the main properties of the $(NX^{--})$ 
bound states, where $N$ stands for a generic BBN-relevant nucleus ($p$, 
$^4$He, $^8$Be, etc.). As is well known from the CBBN studies with singly 
charged particles, the binding energies will be in ${\cal O}($MeV$)$ 
range, {\em i.e.} only marginally smaller than the nuclear binding 
energies relevant for BBN. The binding is achieved via the 
Coulomb force, and for a point-like nucleus, we would have 
$E_{\rm Bohr} = Z_N^2Z_X^2 \alpha^2m_N/2= 2Z_N^2\alpha^2m_N$. (Notice that 
the reduced mass of $X^{--}-N$ system is essentially equal to $m_N$ as 
$m_X\gg m_N$\footnote{Note that 
for a generic DM-nucleus bound state this may not necessarily be true 
\cite{An:2012bs,Berlin:2021zbv}, but here we deal with 
electromagnetically charged relics and very light nuclei participating 
in BBN.}). In practice, however, the 
bound states are never that deep, as the finite radius of the nucleus 
reduces the binding quite considerably, and we resort 
to finding the binding energies numerically. 

We solve the Schr\"{o}dinger equation for 2-body Coulomb problem assuming 
a simplistic nuclear charge distribution. We consider nuclei as 
uniformly charged spheres of radius $R_N = (5/3)^{1/2}r_c(N)$, 
where $r_c$ is the r.m.s.\ charge radius. Direct check against 
{\em e.g.} Gaussian charge distribution reveals the stability of our results 
to ${\cal O}(\%)$ accuracy. 
The binding energies relevant for our study are listed in table~I. These are 
Coulomb binding energies only, {\em i.e.} nuclear polarizability effects are 
not included.

\begin{table}[h!]
  \begin{center}
    \vspace{1cm}
    \label{tab:table1}
    \begin{tabular}{l|c|r} 
    Nucleus $N$ & r.m.s.  charge radius in fm & Binding energy in MeV\\
      \hline
      $p$ & 0.84 & 0.0996 \\
      $^{4}$He & 1.67 & 1.156 \\
       $^{5}$Li & 2.6 & 1.94\\
       $^{6}$Li & 2.6 & 2.10\\
       $^{7}$Li & 2.44 & 2.29\\
       $^{7}$Be & 2.65 & 3.15\\
       $^{8}$Be & 2.5 & 3.40\\
       $^{9}$B & 2.5 & 4.61\\
    \end{tabular}
  \end{center}
  \caption{Coulomb binding energies for $(N\Xmm)$. 
Nuclear r.m.s. charge radii were taken from 
\cite{Angeli:2013epw}, except for highly unstable ${\rm ^5Li}$ and 
${\rm ^8Be}$ nuclei, for which no measurements exist. 
For their charge radii we use extrapolated values. 
}
\end{table}

Examining this table, we can make the following observations: 
\begin{enumerate}

\item Only for proton the actual binding energy is within 1\% of the Bohr 
formula, while for the rest of nuclei the deviation is more than 50\% 
due to the finite charge radius. Importantly, we neglect backreaction of 
$\Xmm$ on nuclear shapes, and assume the nuclei to be not deformed 
inside the bound states. 

\item $^{4}$He is the most abundant nucleus after $p$, and 
it forms earlier than $(p\Xmm)$. Recall that the usual SBBN 
deuterium bottleneck (the onset of nuclear reactions that uptake $n$) is 
at $T\simeq 90$ keV, and it is $\sim 25$ times smaller than 
the deuteron binding energy of 2.2 MeV. Therefore, we can immediately 
conclude that the main $(NX^{--})$ bound states 
will start forming below $T\simeq 40$\,keV. This will lead to significant 
simplifications, because only a handful of reaction are still going, and 
$Y_{\rm He} = const$ at these temperatures.

\item 
Mass five nuclear divide ({\em i.e.} \!the absence of stable 
$A=5$ nuclei) {\em is not bridged} by $\Xmm$. 
Since the rest energy excess of $^{5}$Li compared to $^4$He+$p$ is 1.69 MeV, 
the $({\rm ^5Li}\Xmm)$ binding energy of 1.94 MeV is not enough to 
compensate for that. (The relevant energy is the relative binding energy, 
$1.94-1.15\simeq0.8 $\,MeV, which is smaller than 1.69 MeV). 
Similar considerations and the same conclusion apply also 
for $(^{5}$He$\Xmm)$ bound state. Therefore, mass 5 nuclei do 
not form even in the binding with $\Xmm$, and we can ignore them.

\item Mass eight divide is definitely bridged for $^8$Be and $^9$B, as their 
mass excess is ${\cal O}(0.1\,{\rm MeV})$ while their binding 
energies to $X^{--}$ are around 3 to 5 MeV. We expect that a 
significant fraction of $\Xmm$ will end up in the bound states with 
these elements. One can also check that there is no $\beta^+$ 
or electron capture decay of ($^{9}$B$\Xmm$) to ($^{9}$Be$\Xmm$),  
though $\beta^-$ decay (${\rm ^{9}Be}\Xmm)\to ({\rm ^9B}\Xmm$) is 
still possible.

\item Completely accidentally, our accuracy is not enough to determine 
the relative stability of ($^{7}$Li$\Xmm$) {\em vs.} ($^{7}$Be$\Xmm$). 
In the absence of $\Xmm$, $^7$Be decays to lithium via electron capture, 
but with the calculated bindings 
to it, the total energies of ($^{7}$Be$\Xmm$)+$e$ 
and ($^{7}$Li$\Xmm$) are degenerate within a few keV, whereas we cannot trust 
our methods to that level. If the decay proceeds towards 
($^{7}$Li$\Xmm$), {\em and} the lifetime of $\Xmm$ is comparable to the 
age of the Universe, then ($^{7}$Li$\Xmm$), 
whose charge is +1, will chemically appear, 
after the electron recombination, as an abnormal isotope of hydrogen, while 
in the opposite case, the surviving state will be 
($^{7}$Be$\Xmm$) 
(charge +2), which will be perceived as heavy helium.

\item If $\Xmm$ is stable on the modern Universe timescale, the abundance of 
($^{6}$Li$\Xmm$) will currently appear as abnormal hydrogen. Therefore, 
calculating this yield is also of interest.

\begin{figure}
\centering
\includegraphics[width=10cm]{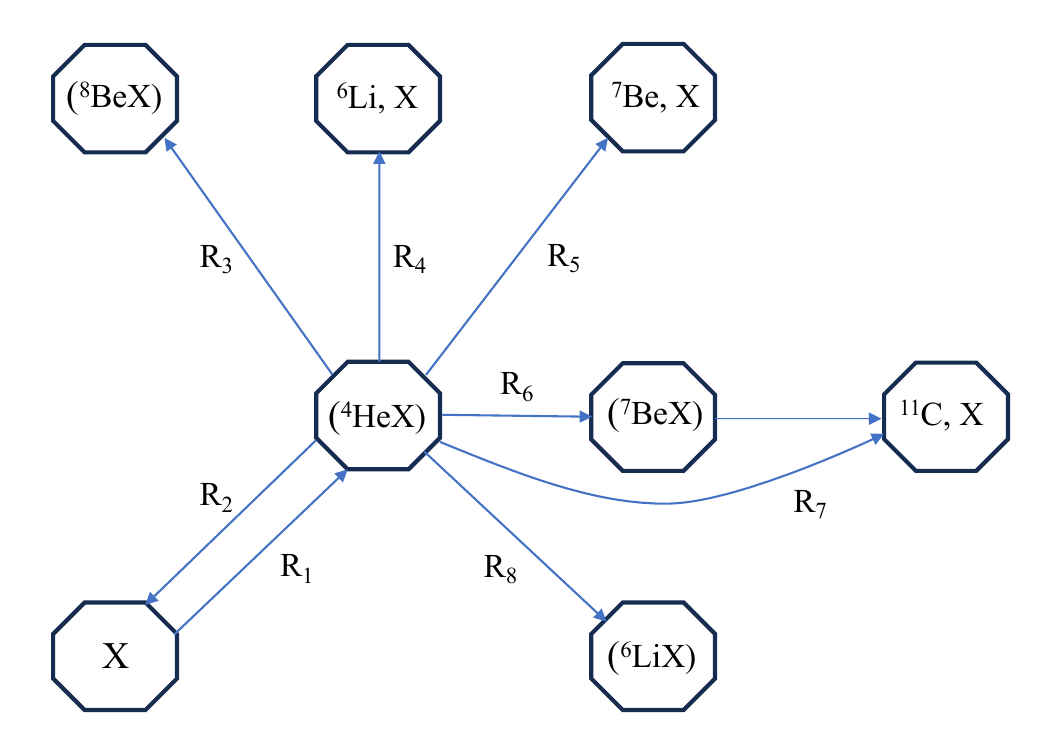}
\caption{Schematic representation of the most relevant CBBN reactions 
with $\Xmm$. Brackets $(NX)$ indicate a bound state of a nucleus $N$ 
with $\Xmm$, while $N,\,X$ stand for a ``deconfining" transition, with 
the final-state $N$ and $\Xmm$ no longer attached to each other. 
}
\label{CBBNf}
\end{figure}

\end{enumerate}

For some ($N\Xmm)$, it turns out, we will also need to calculate the 
properties of their excited atomic bound states. These properties can be found 
again by solving the corresponding Schr\"{o}dinger equations. The results 
should in general be even more reliable than those for the ground states, 
on account of a larger separation between $\Xmm$ and the nucleus, resulting 
in smaller corrections due to the finite nuclear radius. 

Armed with the knowledge of bound states, we can proceed to building the 
catalyzed BBN reaction network and calculating the yields of 
various elements. The key ``element" for us is 
the bound state of $\Xmm$ with $^4$He, as helium is quite abundant, and 
$({\rm ^4He}\Xmm)$ is electrically neutral, so that its reactions 
with other nuclei are not suppressed by Coulomb barriers. 
This makes $({\rm ^4He}\Xmm)$ play a central role in CBBN. 
The schematics of 
main CBBN process is given in Fig.~\ref{CBBNf}. Our goal is to 
calculate or estimate the rates of the main reactions $R_i$ and calculate the 
yields of $^6$Li, $^7$Be+$^7$Li, and of heavier elements, such as $^{11}$B. 
The latter turns out to be mostly initially produced as $^{11}$C in our 
case. 

In the final states of certain reactions ($R_4$, $R_5$, $R_7$), the 
nuclei and $\Xmm$ are no longer attached to each other. These are the reactions 
that are enhanced by the heavy charged particle $X$ the most, as they 
are non-radiative and thus allow  
rates that are much larger than the SBBN rates \cite{Pospelov:2006sc}.

Throughout our calculations, we will treat the $\Xmm$ as a 
perturbation to the main SBBN reaction network, and work in the first 
order in their concentration. 
In other words, we will neglect 
all reactions that involve two or more of $\Xmm$ in the initial state. 
This is justified, in some sense, by our 
results, as the constraints imposed by Li/Be/B produced in CBBN turn 
out to be quite strong, and the resulting maximally allowed abundances 
of $\Xmm$ are indeed very small, especially for long lifetimes.

As we are going to see, the main CBBN processes are delayed until the 
temperatures drops below $40$ keV, when the SBBN reaction are mostly complete: 
apart from a small leftover, all neutrons are already incorporated into 
$^4$He. At this point of the evolution one can identify three groups of 
primordial elements. First, $p$ and $^4$He, that have ${\cal O}(0.1-1)$ 
abundances, then, $n$, D, $^3$He and T, whose abundances are in the range 
${\cal O}(10^{-7}-10^{-5})$. Finally, at ${\cal O}(10^{-10})$ 
and below are Li/Be/B. It is this last group of elements that will be 
affected by CBBN reactions involving $\Xmm$, while the first two groups are not 
changed by $\Xmm$ in our approximation, and their abundances can be used 
as an input from the SBBN reaction network.

In the rest of this section, we consider radiative reactions 
$R_1, R_2$ and $R_3$ that are most important for 
determining the evolution of the \hex\ bound state: 
\begin{eqnarray}
    R_1:&&~~^{4}{\rm He} +\Xmm \to (^{4}{\rm He}\Xmm) + 
\gamma,~~~Q=1.15\,{\rm MeV} 
\label{R1}
\\
    R_2:&&~~(^{4}{\rm He}\Xmm) + \gamma\to {\rm ^4He} +\Xmm, 
~~~Q=-1.15\,{\rm MeV}
\label{R2} \\
    R_3:&&~~(^{4}{\rm He}\Xmm) +{\rm ^4He} \to (^{8}{\rm Be}\Xmm) + 
\gamma,~~~Q=2.16\,{\rm MeV}. 
\label{R3}
\end{eqnarray}
Notice that the $Q$-value of reaction $R_3$ takes into account the 92\,keV 
mass excess of free $^8$Be relative to decay into $2\alpha$. 

For reactions $R_1$ and $R_2$ the calculations closely follow the standard 
calculations of the photoelectric effect in hydrogenic systems.
It is convenient to express the cross section of \hex\ photo-disintegration  
reaction $R_2$ in the following form: 
\begin{equation}
\sigma_{\rm photo} = \frac{2^8Z_{\rm He}^2\alpha\pi^2}{3\exp(4)m_{\rm 
He} \omega} \times \frac{\rho^2_{rad}}{\rho^2_{rad,C} }.
\end{equation}
In this expression, $\omega \simeq Q = 1.15$ MeV is the energy of the 
absorbed $\gamma$, $Z_{\rm He}=2$ is the charge of the particle that 
gets detached, and the last factor is the ratio of radial integral 
$\int r^2dr R_{k1}R'_{10}$ calculated for realistic atomic 
wave functions 
to that for the Coulomb ones. When this ratio is set equal to unity and 
$\omega$ is chosen to be hydrogenic, {\em i.e.} $Z_{\rm He}^2 Z_X^2 \alpha^2 
m_{\rm He}/2$, Eq.~(\ref{R2}) 
reduces to the textbook result for the photoelectric effect at the 
threshold. The cross section of the 
inverse ({\em i.e.}``recombination'') reaction $R_1$ is given by 
$\sigma_{\rm rec} = \sigma_{\rm photo} \times 2 \omega^2/k_{\rm He}^2$, 
where $k_{\rm He}$ is the momentum of the initial-state ${\rm ^4He}$. 
We approach these reactions in a simplified manner, and use hydrogenic wave 
functions, though more elaborate calculations can be performed as 
well. With these cross sections, we find for the rates for reactions $R_1$ 
and $R_2$ 
\begin{eqnarray}
\Gamma_{1}(T_9) &=& \langle\sigma v\rangle_1n_{\rm He}=0.03 \times 
T_9^{5/2} \,{\rm {s}^{-1}},
\label{Gamma1}\\
\Gamma_{2}(T_9) &=&\langle\sigma v\rangle_2n_{\gamma}= 1.88\times 
10^{15} \times T_9 \exp[-13.3/T_9] \,{\rm {s}^{-1}},
\end{eqnarray}
where $T_9$ is the temperature in units of $10^9$\,K ($T_9=1$ is 
equivalent to $T=86.2$\,keV). 
Note that $\Gamma_1$ is the rate of reaction $R_1$ {\em per one $\Xmm$ 
particle}, whereas $\Gamma_2$ is the rate of $R_2$ per \hex\ bound state. 
These are the rates to be carried to the system of Boltzmann equations. 
When $\Gamma_i$'s become larger than the Hubble expansion rate,  
\begin{equation}
H(T_9) = 2.8\times 10^{-3}\times T_9^2 \,{\rm {s}^{-1}}, 
\label{Hubble}
\end{equation}
the corresponding reactions tend to approach equilibrium. It is easy to see 
that $\Gamma_1> H$ for all $T_9<1$, and it is the rapidly falling 
with $T$ rate of $R_2$ that 
determines the survival of \hex. This is, of course, in direct analogy with 
the delayed SBBN deuterium formation, and it is appropriate to call it 
``the \hex\ bottleneck".

Another very important reaction for CBBN is $R_3$, as it removes 
\hex\ and hides $\Xmm$ behind the Coulomb barrier of beryllium, making all 
subsequent reactions with it inefficient. Therefore 
an accurate evaluation of its rate has to be done. 
The overall rate of $R_3$ is given by a combination of the resonant 
and non-resonant parts.  We will use the following strategy for 
calculating the resonant part of this rate, 
corresponding to the channel $(^{4}{\rm He}\Xmm)+{\rm ^4He} \to 
(\rm{^8Be}\Xmm)^*\to (^{8}{\rm Be}\Xmm) + \gamma$. 
The intermediate $({\rm Be}\Xmm)$ in this reaction 
has the excited $n=3$, $l=1$ Coulomb bound state with the 
excitation energy $1.11$ MeV, which is 
almost exactly at the \hex+$^4$He 
separation threshold, with $\Delta E = 0.13$ MeV. (These energy values were 
found numerically by solving the corresponding Schr\"{o}dinger equation.) 
This is a well known narrow resonance situation, when the reaction rate is 
determined by the relative magnitude of the entrance channel width 
$\Gamma_{in}$ and the exit channel width $\Gamma_{out}$. 
The entrance width 
drops out as long as it is larger than the exit 
width but is smaller than temperature. 
For the reaction we consider, this is actually the case. 
Indeed, the entrance width $\Gamma_{in}$ for reaction $R_3$ 
is set by strong and electromagnetic non-radiative processes, 
and we estimate it to be in the 
$\sim 100\,{\rm eV}-{\rm 1\,keV}$ range. 
Thus, it is expected to be much larger than the photon emission widths, which 
for bound states of $\Xmm$ with light nuclei 
can reach ${\cal O}({\rm eV})$. 
The resonant contribution to the 
reaction rate should therefore be determined by the exit rate, which 
is the rate of radiative de-excitation of 
$n=3$ excited Coulomb state of $({\rm ^8Be}\Xmm)$: 
$\Gamma_{out}=\Gamma_{(^{8}{\rm Be}\Xmm),n=3 ~\to~ (^{8}{\rm Be}\Xmm),n=1,2}$. 
This gives 
\begin{eqnarray}
\label{res}
\langle \sigma v \rangle_{3,{\rm res}} = \left( \frac{2\pi}{Tm_{\rm He}} 
\right)^{3/2}\!3\Gamma_{out} \exp[-\Delta E/T]\,\to\,\Gamma_{3,{\rm 
res}}=0.43\times \exp[-1.64/T_9]\times T_9^{3/2} {\rm s}^{-1}.~~~
\end{eqnarray}
In obtaining the last expression we used 
\begin{equation}
\Gamma_{out} = \Gamma_{3p\to 2s} + \Gamma_{3p\to 1s} = \sum_{1s,2s} 
\frac{4}{9}Z_{\rm Be}^2\alpha \omega_\gamma^3 |\langle r \rangle |^2 \simeq 6
\, {\rm eV}\,,
\end{equation}
where the energies $\omega_\gamma$ of the radiative transitions and 
the transition radial matrix 
elements $\langle r \rangle$ were calculated numerically. 

In principle one could also add contributions of other resonances, such as 
$l=2$ and $l=0$ $n=3$ intermediate states, but it turns out that $l=2$ state  
is $\sim 190$\,keV sub-threshold, and for $n=3,l=0$ state the resonance is 
230 keV above threshold, so that we can ignore them both. Also, importantly, 
the inverse reaction can be ignored, as very few photons of 
sufficient energy exist at $T_9<0.5$. We can observe that the resonant rate 
of forming $({\rm ^8Be}\Xmm)$ is fairly large at $T_9\sim 0.3$, but 
it falls off quickly with temperature. 

To estimate the late-time removal of \hex\ via the formation of 
($^8$BeX), we need to know the non-resonant part to the rate of 
$R_3$. Going to extremely slow initial particles, one could argue 
that the $s$-wave scattering should dominate, and therefore the following 
capture reaction must occur,
\begin{equation}
{\rm (^4HeX) +{\rm ^4He} \to ({\rm ^8Be}X)}_{n=2,l=1}+\gamma.
\end{equation}
The $p$-wave final state is uniquely determined by requiring that the 
reaction be exothermic and by the dominance of $E$1 amplitude. The 
{\em intermediate} energy release in this reaction is given by the energy 
differences 
\begin{equation}
\omega_\gamma = Q = (- 1.15 \, {\rm MeV} -(-2.24\, {\rm MeV+0.09\, {\rm 
MeV}} ) ) = +1.00\, {\rm MeV},
\end{equation}
where $2.24$ MeV is the 2$p$ binding energy between $^8$Be and $X^{--}$. 

The calculation of the non-resonant contribution to the rate of $R_3$ is a 
complicated 3-body problem, but what simplifies it is that the interaction 
with photon can be treated as perturbation. The cross section for this 
reaction can be approximated as 
\begin{eqnarray}
\sigma v = \frac{\alpha Z_{\rm He}^2\omega_\gamma^3}{2\pi}
\sum_\lambda \int 
d\Omega_\gamma \left| \int d^3r_1d^3r_2 \psi^*_{{\rm Be}X}\left( 
\frac{\vec{r}_1+\vec{r}_2}{2}\right)\psi^*_{\rm 
Be}(|\vec{r}_1-\vec{r}_2|) \right.\\\nonumber\left.
\times \vec{\epsilon}_\lambda\cdot(\vec{r}_1+\vec{r}_2)\,\psi_{{\rm He}X}
\left(\vec{r}_1\right) \psi_{\rm He}(\vec{r}_2) \right|^2,
\end{eqnarray}
where $\vec{\epsilon}_\lambda$\, is the polarization vector of the outgoing 
photon. The heavy $X$ particle is located at 
the origin, and the coordinates of $\alpha$ particles are $\vec{r}_1$ and 
$\vec{r}_2$. The electric dipole operator in this system is proportional to 
$\vec{r}_1+\vec{r}_2$. This is a 
simplified $\alpha$-cluster picture of $^8$Be as being a bound state 
of two $^4$He nuclei in the relative $s$-wave. $\psi_{\rm He}(\vec{r}_2)$ is 
asymptotically the plane wave $\exp(i \vec{k} \vec{r}_2)$ of the incoming 
$\alpha$-particle with momentum $k\sim \sqrt{TM_{\rm He}}$, which 
corresponds to $k^{-1} \sim 20$\,fm. Thus, one can use the long-wavelength  
approximation for the incoming $^4$He. $\psi_{{\rm He}X}\left( 
\vec{r}_1\right) $ and $\psi_{{\rm Be}X}\left(
\frac{\vec{r}_1+\vec{r}_2}{2}\right)$ are the bound state wave functions 
that we have found numerically, 
taking into account the finite nuclear sizes. 
Many of the remaining integrals can be done analytically, exploiting the 
orbital properties ($s$-wave or $p$-wave) of the bound states. The summation 
over the photon polarizations and integration over its directions is 
also done analytically. The final part of the calculation is performed {\em 
assuming} a certain form of the $2\alpha$ wave function inside $^8$Be, 
which we take to be Gaussian, with the widths reproducing the expected 
charge radius of $^8$Be. We also varied this shape, to make sure we are 
not overly dependent on our assumptions. As a result, we arrived at the 
following estimate for the non-resonant part of the cross section: 
\begin{eqnarray}
\label{range}
\sigma v/c \simeq  (3-9)\times 10^{-31}\,{\rm cm}^2.
\end{eqnarray}
Eventually, to get the {\em conservative} bounds on lifetime and 
abundance of $\Xmm$, we need to take the maximum rate for $R_3$, and 
therefore we adopt the largest value of the cross section in 
(\ref{range}), which yields 
\begin{eqnarray}
\Gamma_{3,\rm non-res} \simeq 0.02\times T_9^3\,{\rm s}^{-1}.
\label{nonres}
\end{eqnarray}
The sum of the (\ref{res}) and (\ref{nonres}) gives our final estimate 
for $\Gamma_3$, which for $T_9\simeq 0.2 - 0.4$ 
is only marginally smaller than $\Gamma_1$. 

\section{Catalyzed reactions}

Once the \hex\ state is formed, it can react in a variety of ways. As 
already mentioned, $R_3$ removes some of these states from the play and 
hides them inside ($^{8}$Be$\Xmm$) that can further transform to 
($^{9}$B$\Xmm$). It is interesting that if $\Xmm$ is unstable, these 
extra bound states are relatively harmless, as outside the bound states 
they immediately decay to $2\alpha$ and $p2\alpha$.  It has to be 
mentioned that the reaction $(^8{\rm Be}\Xmm)+n \to {\rm^9Be} + \Xmm$ 
has a negative $Q$-value and can be neglected. This is in contrast with 
CBBN with singly charged particles, where such reaction is catalyzed 
\cite{Pospelov:2007js,Pospelov:2008ta}. The rate for the $(n,\gamma)$ 
reaction on $(^8$Be$\Xmm)$ will be of course additionally suppressed. 
Therefore, the reactions of \hex\ with other elements, pictured 
in Fig.~\ref{CBBNf}, that are of most interest to us, are 
\begin{eqnarray}
R_4:&&~~(^{4}{\rm He}\Xmm) +{\rm D} \to \Xmm + {\rm ^6Li},~~~Q=0.31\,{\rm MeV} 
\label{R4}\\
R_5:&&~~(^{4}{\rm He}\Xmm) +{\rm ^3He} \to \Xmm + {\rm ^7Be},~~~Q=0.41\,
{\rm MeV}\label{R5}\\
R_6:&&~~(^{4}{\rm He}\Xmm) +{\rm ^3He} \to ({\rm ^7Be}\Xmm)+\gamma,~~~Q=3.56\,
{\rm MeV}\label{R6}\\
R_7:&&~~(^{4}{\rm He}\Xmm) +{\rm ^7Be} \to {\rm ^{11}C}+\Xmm,~~~Q=6.39\,
{\rm MeV}\label{R7}\\
R_8:&&~~(^{4}{\rm He}\Xmm) +{\rm D} \to ({\rm ^6Li}\Xmm)+\gamma,~~~Q=2.42\,
{\rm MeV}.
\label{R8}
\end{eqnarray}
We could also include other reactions, {\em e.g.} with $^3$H, but by the 
time \hex\ is formed, tritium concentration is already considerably depleted.

Before performing any calculations, we note that the rates for the first 
two reactions here dominate, $\Gamma_{4,5} \gg \Gamma_{6,7,8}$. Reaction 
$R_7$ is catalyzed, but its rate is suppressed due to the small SBBN 
abundance of $^7$Be, while $R_{6,8}$ 
are radiative, which leads to suppression of their rates. 
We still need reaction $R_8$ for 
the following reason: for a very long-lived $\Xmm$, $\tau_X>13\,{\rm 
bn\,yr}$, its bound state with $^6$Li will 
appear as a super-heavy isotope of hydrogen.

In order to get estimates for the rates of $R_{4,5}$, we use the 
previously calculated rates for reactions with singly charged $X^-$. In 
particular, Ref.~\cite{Kamimura:2008fx} evaluates $(^4{\rm He}X^-)+{\rm D}\to 
{\rm ^6Li}+X^-$ and $(^4{\rm He}X^-)+{\rm ^3He}\to {\rm ^7Be}+X^-$, finding 
the astrophysical $S$-factors for both. We make the assumption that the 
$S$-factors remain roughly the same for $\Xmm$ and, taking into account 
that $(^4{\rm He}\Xmm)$ is electrically neutral, rescale the rates 
found in~\cite{Kamimura:2008fx} by ``stripping off" the Coulomb barrier 
penetration factor ${\cal P}$. 
This factor can be written as 
\begin{equation}
{\cal P} = 
\sqrt{\frac{E_G}{E}}\exp\left(-\sqrt{\frac{E_G}{E}}\right), 
\end{equation}
where $E_G$ 
is the Gamow energy for the reaction with singly-charged $X^-$, 
$E_G = 2\pi^2\alpha^2Z_N^2 m_N$. Here $m_N$ is 
the mass of the incoming nucleus, $m_{\rm D}$ for reaction (\ref{R4}) and 
$m_{\rm ^3He}$ for reaction (\ref{R5}). 

We also need to take into account Sommerfeld/Coulomb enhancement in the 
final state, described by the factor 
\begin{equation}
{\cal C}(Z_X) = \frac{2\pi Z_X Z_N \alpha}{v}.
\end{equation}
This quantity is actually $Q$-value dependent, as $v=\sqrt{2Q/m_N}$.  
To properly rescale the results of \cite{Kamimura:2008fx}, we need to 
multiply them by ${\cal C}(Z_X=2)/{\cal C}(Z_X=1)$. 
Numerically, we find that for reactions $R_4$ and $R_5$ these rescaling 
factors are 3.9 and 3.5, respectively. 

Putting it all together, one gets the following approximate formula: 
\begin{equation}
\langle \sigma v \rangle_{4(5)} = \frac{{\cal C}(2)}{{\cal 
C}(1)}\frac{S_{4(5)}}{\sqrt{E_Gm_N/2}}\,.
\end{equation}
This expression is velocity-independent in the limit of small velocity, 
as it should be. The $S$-factors here correspond to catalysis with singly 
charged $X^-$; following  
\cite{Kamimura:2008fx}, we take $S$ to be 
$\simeq 45$ b\,keV for reaction $R_4$ and $\simeq 15$ b\,keV for reaction 
$R_5$.

We can now evaluate the corresponding quantities and reaction rates. In the 
standard notation for the reaction rates, we obtain the following 
expressions:
\begin{eqnarray}
\lambda(R_4) \equiv N_A(\sigma v)_4 = 6.9 \times 10^7
{\rm ~cm^3\,s^{-1} mole^{-1}},~~
\label{lambda4}\\
\lambda(R_5) \equiv N_A(\sigma v)_5 = 7.3 \times 10^6
{\rm ~cm^3\,s^{-1} mole^{-1}},~~ 
\label{lambda5}
\end{eqnarray}
where $N_A$ is the Avogadro constant. 
This gives the following rates per \hex\ to be used in the Boltzmann BBN 
network:
\begin{align}
&\Gamma_4 = (\sigma v)_4\,n_{\rm D} = 0.03 \times T_9^3{\rm ~s^{-1}},~~
\label{Gamma4} \\
&\Gamma_5 =(\sigma v)_5 \,n_{\rm ^3He} = 1.1\times 10^{-3} \times T_9^3
{\rm ~s^{-1}}.
\label{Gamma5} 
\end{align}
Here we used the SBBN abundances of D and $^3$He. (In fact, 
$\Gamma_4$ gets additional mild evolution with temperature, as D/H is still a 
slow-varying function of temperature below $T_9=0.4$.) Very importantly, 
$\Gamma_5/H$ stays above 1 until $T_9 =0.08$, which guarantees very 
large output of $^6$Li per $\Xmm$.

We can further assess the rates for other reactions of interest. 
Consider first reaction $R_6$ (\ref{R6}). If $\Xmm$ is stable on the 
timescale of the age of the Universe, and if $(^7$Be$\Xmm)$ does not decay to 
$(^7$Li$\Xmm)$, it should currently appear as a heavy isotope of He, just as 
$(^8$Be$\Xmm)$ does. However, its abundance is much smaller than that of 
$(^8$Be$\Xmm)$, hence reaction $R_6$ can be safely ignored in this 
case. If $\Xmm$ is unstable, then, after the decay of $\Xmm$, $(^7$Be$\Xmm)$ 
gives a contribution to $^7$Be, and ultimately to $^7$Li. The rate of 
$R_6$ can therefore be added to that of $R_5$ and, since
$\Gamma_6\ll \Gamma_5$, we do not consider it further. 

The rate of reaction $R_7$ is not known. However, we can make a crude estimate 
for this rate by assuming that $^7$Be and $^4$He coalesce into ${\rm ^{11}C}$ 
with the cross section $\sim \pi (R_{\rm He}+ R_{\rm Be})^2 
v_0/v $, where $v_0\sim Z_{\rm He}\alpha$ is the characteristic velocity 
of helium on its orbit in 
$({\rm ^4He}\Xmm)$ and $v$ is the relative velocity of the colliding 
$({\rm ^4He}\Xmm)$ and 
${\rm ^7Be}$. This yields 
\begin{eqnarray}
    (\sigma v)_7 \sim  2.5\times 10^{-26}\times c~~~\to ~~~\Gamma_7 = 
  (\sigma v)_7\times n_{\rm Be_7} = 3.5\times 10^{-6}\,T_9^3\,{\rm s}^{-1}.
\end{eqnarray}
Here $n_{\rm Be_7}$ stands for the SBBN number density of ${\rm ^7Be}$. 
Notice that $\Gamma_7/H \ll 1$ at all times due to the extremely small 
SBBN abundance of $^7$Be. Therefore, the output of $^{11}$C per $\Xmm$ 
will remain small. The production of ${\rm ^{11}C}$ will, however, have the 
following interesting consequence. 
$^{11}$C is virtually stable at the BBN temperatures, and it has a 
delayed decay to $^{11}$B. Therefore, a significant fraction of newly 
created $^{11}$B will be shielded from the proton-induced burning. 
Weak decay of $^{11}$C and reaction $^{11}$B+$p\to 3\alpha$ 
are added to our CBBN reaction network with the standard values of their rates 
(see {\em e.g.} \cite{Caughlan:1987qf}).

Finally, the rate of reaction $R_8$ that determines the abundance of stable 
$(^6{\rm Li}\Xmm)$ when $\Xmm$ is stable on the timescale of~$\gtrsim$ 
13 bn years can be estimated from the following considerations. 
While this is a photon emission rate and therefore it is several orders of 
magnitude below that of reaction $R_4$, it is still much enhanced compared 
with the rate of the SBBN reaction $^4{\rm He}+{\rm D}\to{\rm ^6Li} + \gamma$. 
The reason is that the \hex+D system does have an unsuppressed $E1$ dipole 
operator $\vec{d} = 
Z_{\rm D}e \times \vec{r}_{\rm D}$, and therefore one should expect this 
reaction to have a Coulomb-unsuppressed rate characteristic for $E1$ 
transitions. More specifically, we expect that the replacement of $^4$He 
by \hex\ amounts to the enhancement of the $E1$ amplitudes by the factor 
\begin{eqnarray}
    {\cal S}=\frac{|\vec{d}_{\rm (^4He\Xmm)-D}|}{|\vec{d}_{\rm 
^4He-D}|}=\frac{m_\alpha+m_{\rm D}}{m_{\rm D}m_\alpha}\times 
\left(\frac{2}{m_\alpha}-\frac{1}{m_{\rm D}} \right)^{-1} \simeq 230\,.
\end{eqnarray}
To estimate the rate for $R_8$, we use the calculations of $E1$-induced 
transition in the standard $^4$He+D fusion, Ref. \cite{Ryzhikh:1995zz}, 
that gives $\sigma_{\rm D(\alpha,\gamma)^6Li}$ 
on the order of $0.5\times 10^{-33}\,{\rm cm}^2$ at 1.5\,MeV. We then apply 
the following rescaling, taking into account that the cross section for 
$R_8$ scales as $v^{-1}$ in the limit of small velocities:
\begin{eqnarray}
    (\sigma v)_8 = {\cal S}^2 \times (\sigma_{\rm D(\alpha,\gamma)^6Li} 
v)_{E=\rm 1.5\,MeV} \sim 1\times 10^{-30}\,{\rm cm}^2\times c.
\end{eqnarray}
We note that this is quite a typical size for the Coulomb-unsuppressed 
cross section with $E1$ photon emission. This way we arrive at the 
reaction rate for $R_8$,  
\begin{eqnarray}
    \Gamma_8 = (\sigma v)_8 \times n_{\rm D} \simeq 8\times 
    10^{-6}\, T_9^3\,{\rm s}^{-1},
    \label{R8total}
\end{eqnarray}
where we used the SBBN value for deuterium concentration $n_{\rm D}$. As 
expected, this rate is below the Hubble expansion rate.

\section{CBBN yields of Li/Be/B} 

Having determined the most relevant catalyzed rates, we can build the CBBN 
reaction network, assuming $\Xmm$ to be a small linear perturbation, and 
following the fate of the bound states and the catalyzed Li/Be/B production.

The evolution of the bound states of $\Xmm$ is shown in 
Fig.~\ref{BoundStates}. The process of their formation indeed starts at 
$T_9<0.4$, and most doubly charged particles end up inside these bound states. 
With our estimate of the non-resonant part of $\Gamma_3$, we conclude 
that most (about 80\%) of $\Xmm$ end up inside $({\rm ^8Be}\Xmm)$, while only 
about 20\% are preserved in the form of highly reactive \hex. Should we 
completely neglect the non-resonant part of $R_3$, the balance would shift 
 towards \hex, as it reaches 90\% in that case. Still, this uncertainty is 
acceptable, as it shows that indeed under any assumptions large 
quantities of \hex\ per $\Xmm$ are preserved.

It is instructive to discuss also the subsequent fate of the free 
$\Xmm$. At temperatures below $T_9 \sim 0.03$ they will start forming 
bound states with protons, $p+\Xmm \to (p\Xmm)+\gamma$. This is 
followed by the extremely fast charge exchange reaction with helium (as 
was investigated for singly charged relics in 
\cite{Pospelov:2008ta}), $(p\Xmm) + {\rm ^4He} \to$\hex$+p$. This 
completely eliminates the population of free $\Xmm$, but is not relevant 
for our discussion as it has only a minor effect on the \hex\ abundance.

\begin{figure}
\centering
\includegraphics[width=10cm]{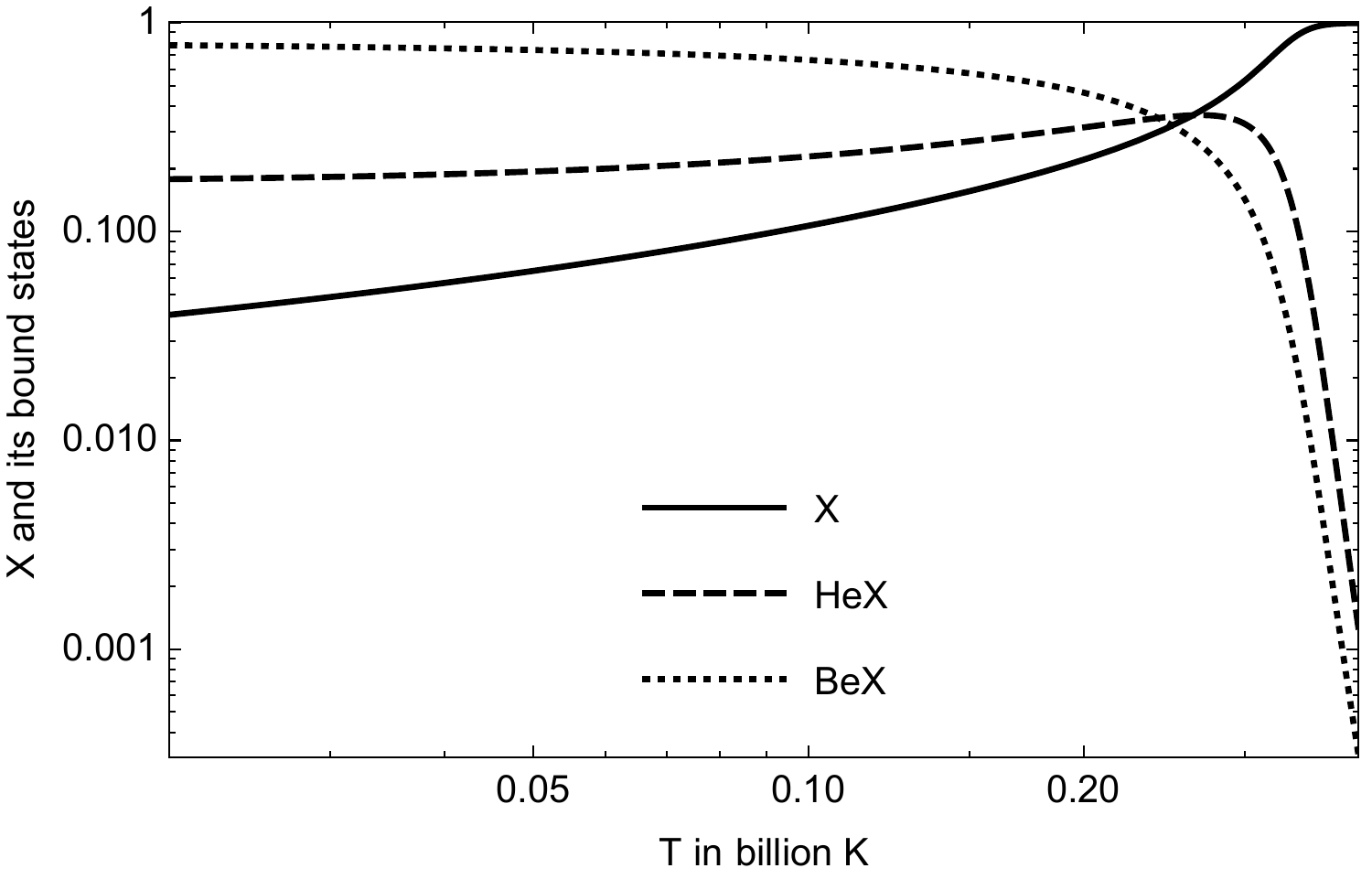}
\caption{Evolution of $\Xmm$ and of its bound states in the limit of 
parametrically long lifetime. The sum of all concentrations here is 
equal to one by definition. The graph shows that the main uptake of 
$\Xmm$ happens below $T=30$\,keV, and ($^8$Be$\Xmm$) eventually absorbs 
about 80\% of doubly negatively charged particles. 20\% remain in the 
most reactive form \hex. The concentration of free $\Xmm$ drops to a 
few percent level mark. }
\label{BoundStates}
\end{figure}

Next, we investigate the yields of Li/Be/B, upon the inclusion of all 
CBBN rates. At this point we will keep the lifetime of $\Xmm$ as a 
variable, and analyze the outcome as a function of $\tau_X$. 
Fig.~\ref{tauinf} shows the evolution curves for the main elements 
we are interested in. As is clearly visible in the figure, the abundance of 
$^6$Li is initially rather suppressed because of the very efficient 
$(p,\alpha)$ burning of this element. However, at $T_9\sim 
0.1$ this burning stops, and the yield of $^6$Li comes to dominate the 
Li/Be/B BBN predictions. Boron and $^7$Be in that sense are relatively 
suppressed. Despite the fact that the formation of $^6$Li is 
Coulomb-unsuppressed, at certain point it still decouples, simply 
because the corresponding rate, weighted by the Hubble rate, drops with 
temperature, $\Gamma_4/H\sim T_9$. The small temperature limit of the curves 
in Fig.~\ref{tauinf} allows us to determine the yields of the relevant 
elements per $\Xmm$ particle. Thus, for parametrically long 
lifetimes $\tau_X\gg \tau_{\rm BBN}$ we get the following freeze-out 
yields:
\begin{eqnarray}
\label{CBBNlargetau}
\frac{{\rm ^6Li}}{X^{--}} \simeq 0.25;~~~
\frac{{\rm ^7Be}}{X^{--}} \simeq 0.04;~~~
\frac{{\rm ^{11}B}}{X^{--}}\simeq 0.7\times 10^{-4}.
\end{eqnarray}
Now we can compare these predictions with the observational limits on 
the $^6$Li abundance, Eq.~(\ref{limit_Sun}), and we immediately conclude that 
$\Xmm/{\rm H}$ must satisfy   
\begin{eqnarray}
   \left. \frac{X^{--}}{\rm H} \right|_{\tau_X\gg \tau_{\rm BBN}} < 10^{-9}.
   \label{finallimit}
\end{eqnarray}
This is a much stronger limit, by about four orders of 
magnitude, than the one previously found for $X^-$, which is 
mostly due to the absence of the Coulomb barrier for \hex.

One example of the yields of Li/Be/B for a finite lifetime of the $\Xmm$ 
particles is given in Fig.~\ref{tauinf}, lower panel, where the 
lifetime is chosen to be 500 seconds. In that case the 
production of Li/Be/B stops early, as there is no late time abundance of 
$X$ particles, and nearly all $^6$Li is burnt. However, $^7$Be is 
preserved, and its yield is $\sim 10^{-3}$ per $\Xmm$ particle. 
Using the conservative upper limit on $^7$Li production 
in BBN (\ref{limit_Sun_Li7}) will limit the initial abundance of $\Xmm$ to be 
less than $10^{-5}$. (The initial abundance here is defined as 
$(\Xmm/{\rm H})|_{t\ll t_{\rm BBN}}$).

\begin{figure}
\centering
\includegraphics[width=9.5cm]{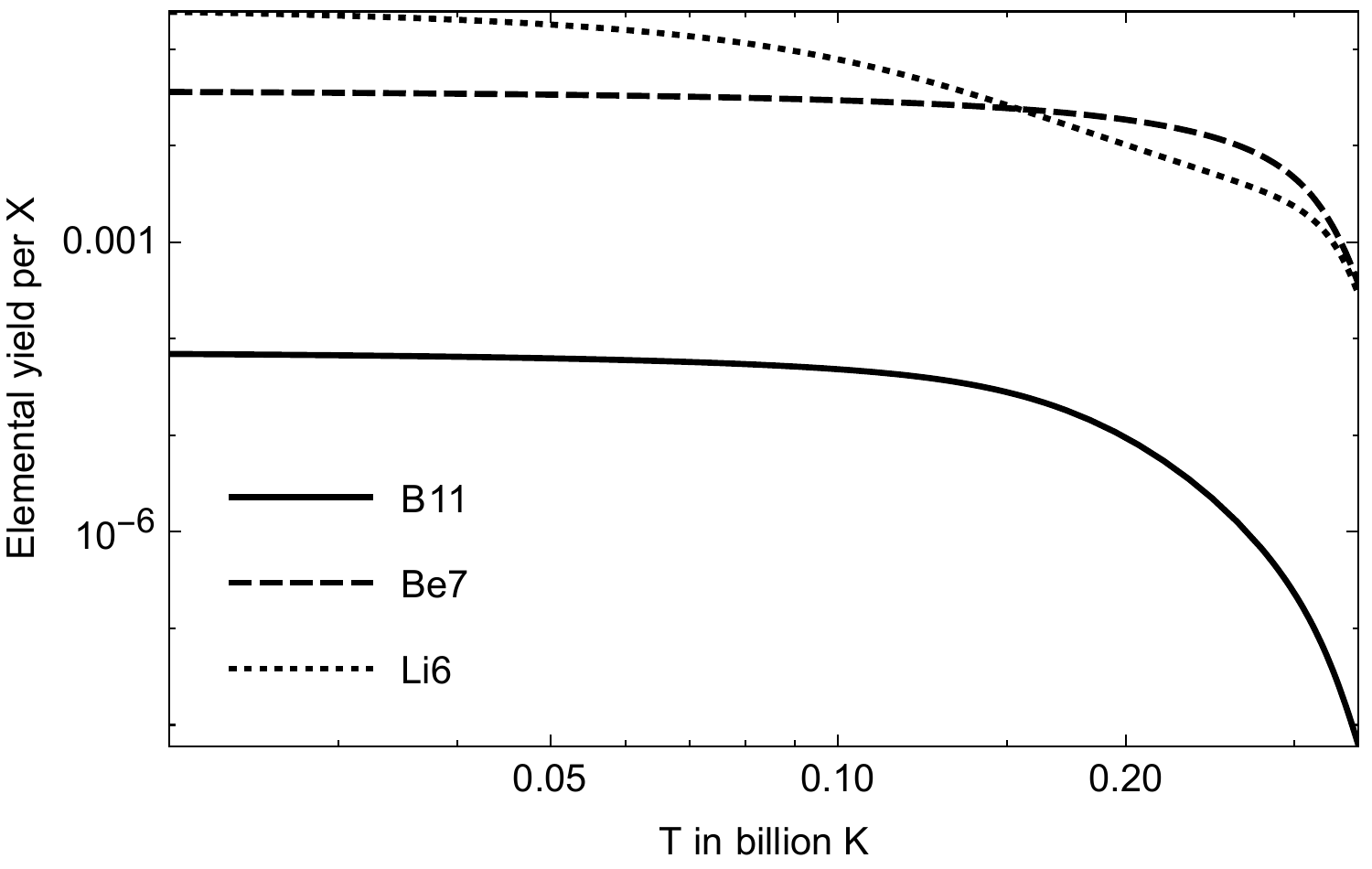}
\includegraphics[width=9.5cm]{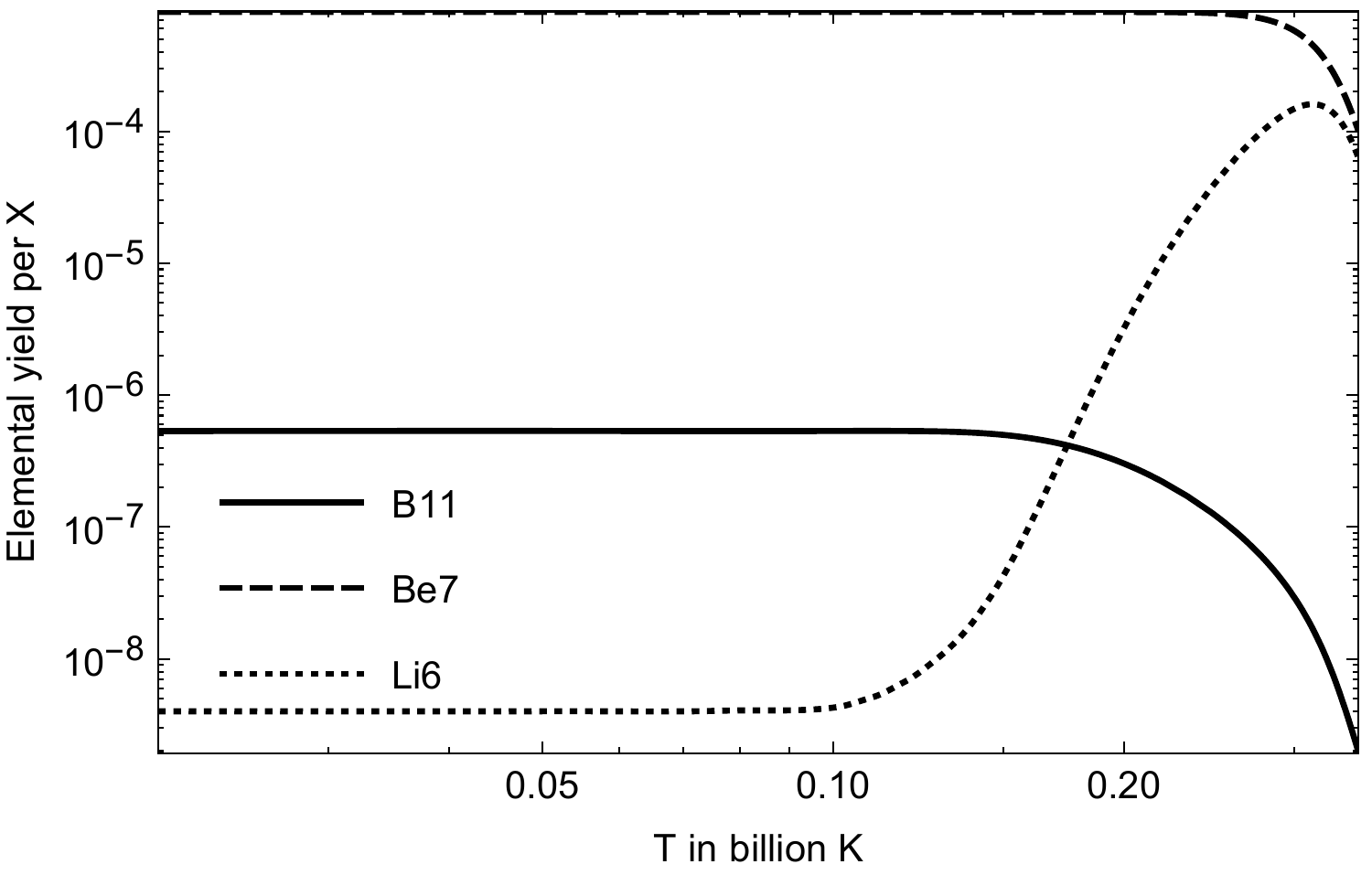}
\caption{{\em Upper panel:} Yields of Li/Be/B for very long lifetime of 
$X^{--}$. The left ends of the curves give the freeze-out values for the 
elements, {\em per $\Xmm$ particle}. As expected, $^6$Li dominates the 
exotic BBN yields. {\em Lower panel:} Same but for the finite lifetime, 
$\tau_X =500\,{\rm s}$. The suppression of $^6$Li is apparent, and the 
main result of the catalysis is production of $^7$Be$\to^7$Li.}
\label{tauinf}
\end{figure}

Generalizing this to arbitrary lifetimes of the $\Xmm$ particles, we 
derive the constraints on their initial abundance. 
The upper panel of Fig.~\ref{tau-Y} gives the yields of $^6$Li/H and 
$^7$Be/H.  Taking into account that free $^7$Be becomes $^7$Li and making 
use of the conservative limits on $^6$Li and $^7$Li from section 2, 
Eqs.~(\ref{limit_Sun}) and~(\ref{limit_Sun_Li7}), we plot the resulting 
limits on the $\Xmm$/H abundance in the lower panel of the same figure.  As 
expected, the constraint loses its power for short lifetimes, 
$\tau_X <100\,{\rm s}$. For lifetimes $\tau_X< 2500$\,s,  
$^7$Li data have more constraining power, while for longer lifetimes $^6$Li 
is more important.
 
\begin{figure}
\centering
\includegraphics[width=9.5cm]{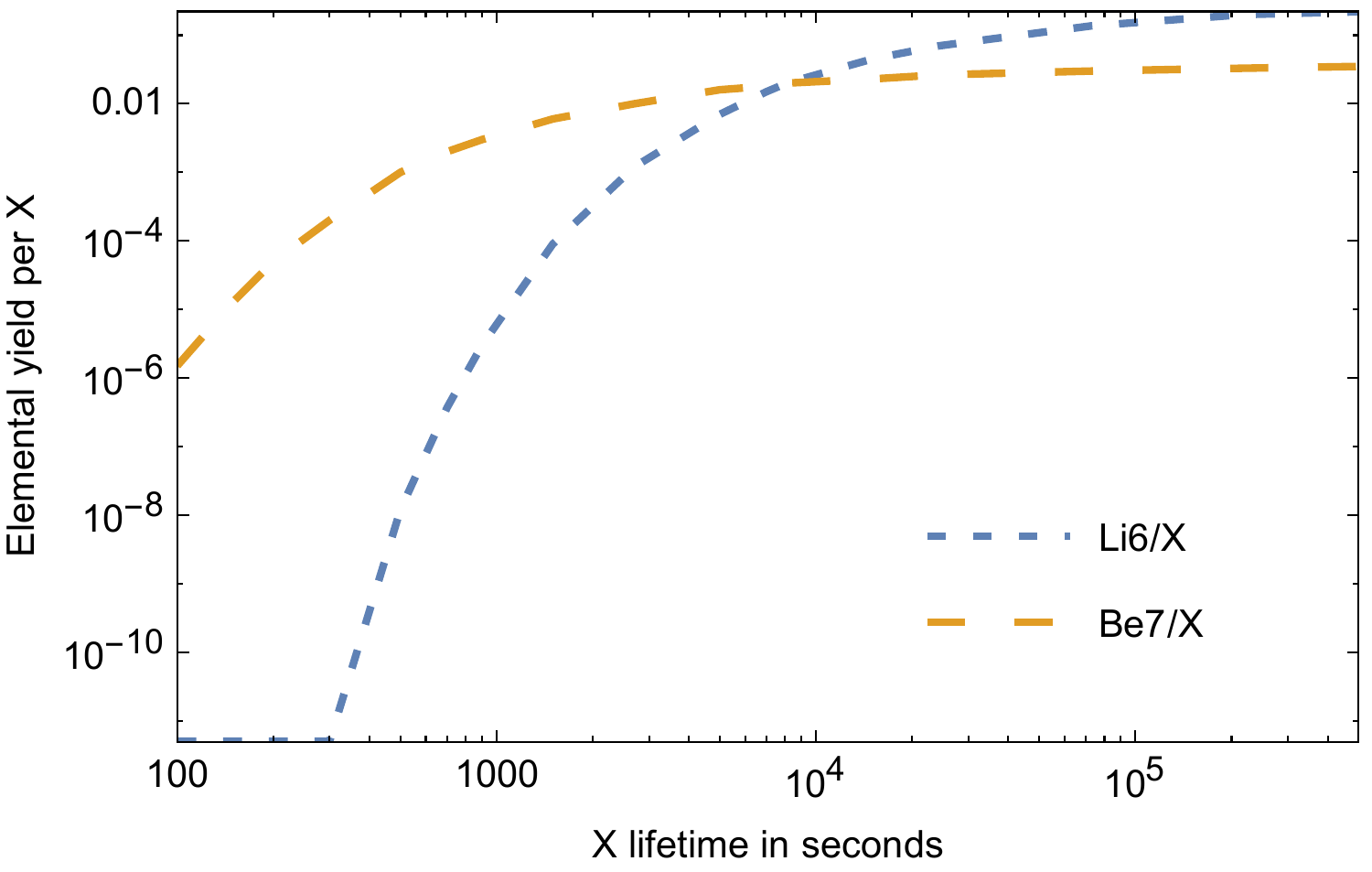}
\includegraphics[width=9.5cm]{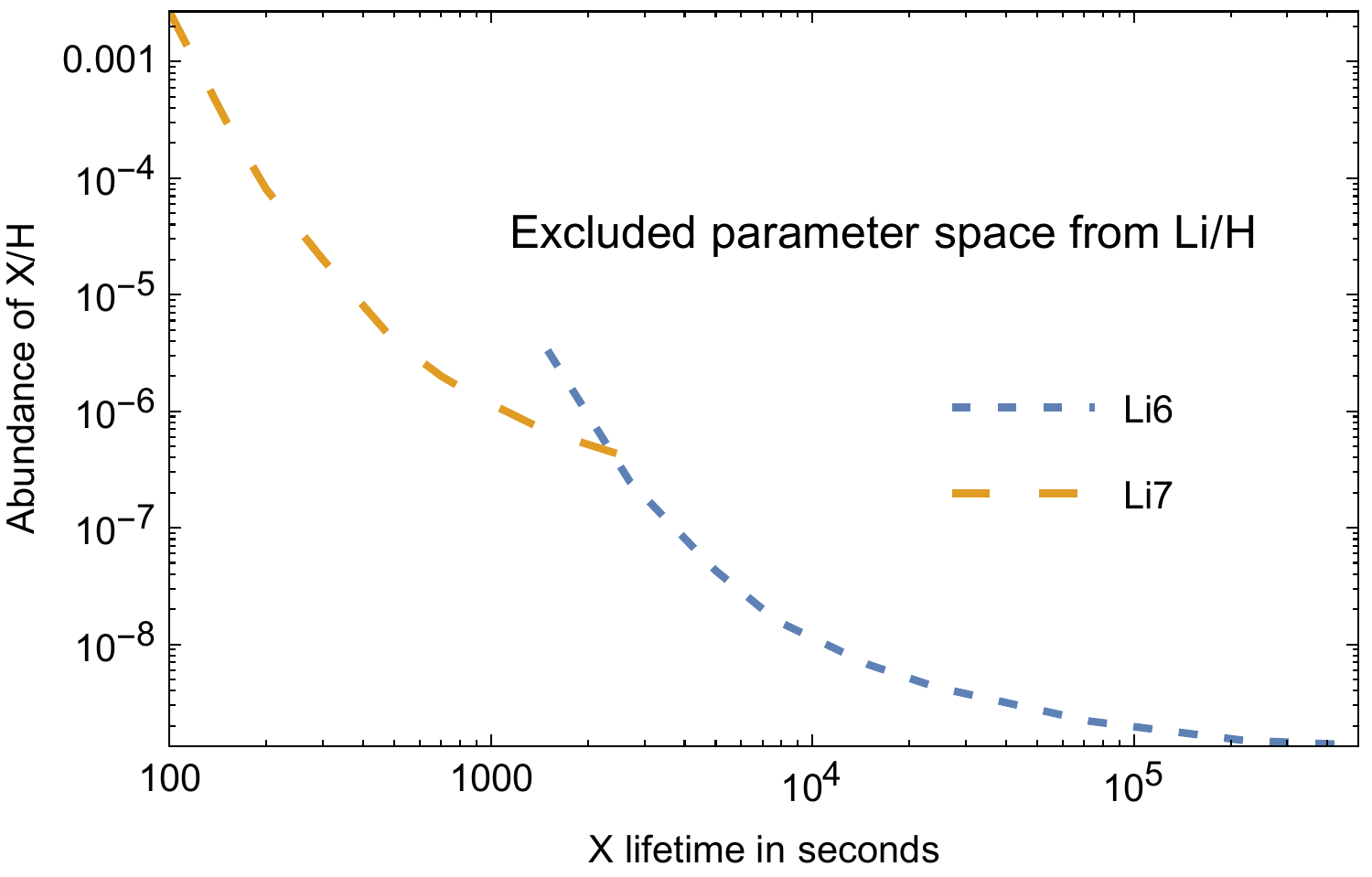}
\caption{{\em Upper panel:} CBBN production of $^6$Li and $^7$Be as a 
function of the lifetime of $X^{--}$. $^6$Li yield is larger by a factor of 
a few at long lifetimes. At short lifetimes, the yield of $^7$Be 
significantly exceeds that of $^6$Li. {\em Lower panel:} Constraints from 
observations of $^7$Li and $^6$Li 
(Eqs.~(\ref{limit_Sun_Li7} ) and (\ref{limit_Sun})) 
on the initial abundance $\Xmm/$H.}
\label{tau-Y}
\end{figure}

\section{Constraints from searches for heavy isotopes of H and He}

If $X$-particles are stable or practically stable ({\em i.e.} if their 
lifetime is comparable to or larger than the age of the Universe), 
doubly positively charged $X^{++}$ would currently appear as an anomalously 
heavy isotope of He. Bound state of $(^8{\rm Be}\Xmm)$ should chemically 
behave in exactly the same way. We predicted, Fig.~\ref{BoundStates}, that 
the relative abundance of these states will be significant, $(^8{\rm 
Be}\Xmm)/\Xmm\sim {\cal O}(1)$. We also found that a small fraction of 
${\rm ^6Li}$ (and possibly ${\rm ^7Li}$) will end up in bound states with 
$\Xmm$. Such states would currently appear chemically as anomalously heavy 
isotopes of hydrogen.

Bound states of the $X$-particles can come to the Earth with CR, or be 
present at the time of  formation of the Earth. Therefore, their abundances 
should be subject to constraints coming from terrestrial searches of 
abnormally heavy isotopes of He and H.

Search for anomalously heavy helium in the Earth's atmosphere was 
carried out using the laser spectroscopy technique in~\cite{Mueller:2003ji}, 
and the limit $10^{-12} - 10^{-17}$ per atom over the mass range 
20 - $10^4$ GeV was obtained. In the scenario with doubly charged 
$X$-particles, it can be directly translated into the 
corresponding limit on the abundance of $\Xmm$:%
\footnote{We note that 
doubly charged long-lived particles with mass $m_X<100$ GeV 
are definitely excluded on account of various collider constraints.}
\begin{eqnarray}
  \left. \frac{{\rm (^8Be\Xmm)}}{\rm H} \right|_{\tau_X> \tau_{\rm 
Universe}} \sim \left.\frac{{ \Xmm}}{\rm H} \right|_{\tau_X> \tau_{\rm 
Universe}} < 10^{-12} - 10^{-17} 
\qquad(10^2\;{\rm GeV}<m_X < 10^4\;{\rm GeV})\,.
  \label{He_isotope}
\end{eqnarray}
This constraint is manifestly much stronger than any CBBN bounds one 
may derive, but, unlike the latter, it is applicable only in the 
$100-10^4$\,GeV mass window. 

A comment on the upper border of this mass interval is in order. 
The study in~\cite{Mueller:2003ji} exploited the isotopic shift of the 
atomic lines and high volatility of normal helium as opposed to much 
reduced volatility of anomalously heavy isotopes. One can expect that the 
constraint should apply as long the mass is not larger than a critical one: 
at very large $m_X$, the gravitational settling velocity of anomalous helium 
will be larger than the typical convective velocities in the atmosphere. 
Due to large atomic-size cross sections, such mass 
may reach rather high values. Ref. \cite{Mueller:2003ji} believes that above 
$m_X>10^4$\,GeV, the efficiency of heavy helium extraction drops, and the 
limits will become uncertain.

As mentioned above, the existence of long-lived $\Xmm$ will also imply the 
existence of abnormally heavy hydrogen isotopes, (Li$\Xmm$). We have already 
pointed out that within the accuracy of our approach it is not possible to 
determine with certainty whether  $(^7{\rm Li}\Xmm)$ or $(^7{\rm Be}\Xmm)$ 
is stable, due to their accidental near degeneracy. If $(^7{\rm Be}\Xmm)$ is 
stable, then $(^7{\rm Li}\Xmm)$ will beta-decay to it, contributing to 
anomalous helium. In the opposite case, $(^7{\rm Be}\Xmm)$ will undergo a 
weak electron capture to $(^7{\rm Li}\Xmm)$, becoming chemically equivalent 
to anomalous hydrogen. Due to this uncertainty, we resort to estimating the 
abundance of ($^6$Li$\Xmm$), which will remain stable once the temperature 
drops to $T_9< 0.1$.

The bound state 
($^6$Li$\Xmm$) can be generated via several pathways. One of them is 
reaction $R_8$, Eq.~(\ref{R8}).  Another pathway involves the charge exchange 
reaction 
\begin{eqnarray}
    ({\rm ^4He}\Xmm)+{\rm ^6Li}\to ({\rm ^6Li}\Xmm)+{\rm ^4He}\,.
\end{eqnarray}
While the cross section of this reaction is very large 
($\sim$geometric), the overall abundance of ${\rm ^6Li}$ is tiny, and we 
expect the (D,$\gamma$) reaction (\ref{R8}) to dominate.

Using (\ref{R8total}), we estimate the yield of $(^6{\rm Li}\Xmm)$ to be
\begin{eqnarray}
  \left. \frac{{\rm (^6Li\Xmm)}}{X^{--}} \right|_{\tau_X
  > \tau_{\rm Universe}} \simeq \frac{\rm (^4He\Xmm)}{X^{--}}\times 
\int_0^{T_9=0.1} \frac{\Gamma_8\,dT_9}{T_9 H(T_9)} \simeq 5\times 10^{-5}.  
  \label{abnH}
\end{eqnarray} 
In Ref.~\cite{Smith:1982qu} search for 
abnormal hydrogen isotopes in water was carried out using 
mass spectrometry methods, and the upper bound of $10^{-28} - 10^{-29}$ 
per nucleon in the mass range $m_X=12 - 10^3$ GeV was established. 
Together with the estimate (\ref{abnH}), it implies 
\begin{equation}
    \left.\frac{{ \Xmm}}{\rm H} \right|_{\tau_X> \tau_{\rm Universe}}  
    < 10^{-23}-10^{-24}
\qquad(12\;{\rm GeV}<m_X < 10^3\;{\rm GeV})\,.
    \label{H_isotope}
\end{equation}
Note, however, that this constraint is applicable in a rather narrow interval 
of $m_X$, a significant portion of which has now been excluded by collider 
data.  

In Ref.~\cite{Verkerk:1991jf} search was performed for abnormally heavy 
hydrogen isotopes in sea water, using centrifugation followed by atomic 
spectroscopy technique. The upper limit of $6\times 10^{-15}$ relative to 
normal hydrogen was obtained for the mass interval $10^4 - 10^8$ GeV. 
Together with (\ref{abnH}), this gives 
\begin{equation}
    \left.\frac{{ \Xmm}}{\rm H} \right|_{\tau_X> \tau_{\rm Universe}}  
    < 10^{-11}
\qquad(10^4\;{\rm GeV}<m_X < 10^8\;{\rm GeV})\,.
    \label{H_isotope2}
\end{equation}
Though this constraint is significantly weaker than that in 
Eq.~(\ref{H_isotope}), it is applicable in a substantially wider interval of 
$m_X$. 

While the limit in Eq.~(\ref{H_isotope}) is demonstratively stronger 
than (\ref{He_isotope}) (within its range of sensitivity to $m_X$), 
one has to keep in mind that it is at the same time less certain; 
the same also applies to constraint (\ref{H_isotope2}). 
Indeed, while abnormal helium is chemically inert, abnormal hydrogen 
isotopes will have a lot more complicated chemistry due to the presence 
of the valence electron. Therefore, while less stringent than 
(\ref{H_isotope}), Eq.~(\ref{He_isotope}) is more robust against 
chemical evolution alteration of terrestrial abundance of (Li$\Xmm$). It 
is also not clear how both limits should be interpreted for masses heavier 
than $\sim1-10$\,TeV, as several assumptions built into these constraints 
would not hold. 

For an extensive review and discussion of non-collider searches of heavy 
stable  particles see Ref.~\cite{Burdin:2014xma} and references therein.

\section{Discussion and conclusions}

We have derived the yields of Li/Be/B from the presence of $\Xmm$ 
particles during the BBN. Our results should be considered rather 
conservative, as we took very relaxed observational upper bounds on the 
maximum yields of $^6$Li and $^7$Li during BBN. The accuracy of our 
predictions is not believed to be better than a factor of a few, however, 
as they partly rely on the rescaling of the catalyzed nuclear rates for singly 
charged particles. In order to achieve a 
more accurate result, one may want to repeat the corresponding calculations 
for $\Xmm$ rather than $X^-$. However, for the log-log plot of 
Fig.~\ref{tau-Y}, these uncertainties are tolerable, and they cannot change 
the overall picture of a huge enhancement of the $^6$Li and $^7$Be yields in 
CBBN. The strength of the CBBN limit at asymptotically long lifetimes, 
$\Xmm/{\rm H} < 10^{-9}$, allows to conclude that the energy injection during 
the CMB  epoch due to {\em e.g.} continuing $R_3$ reaction 
is not going to provide additional constraints. Indeed, the best CMB 
sensitivity is at the level of ${\cal O}(10^{-2}\,{\rm eV})$ of energy 
injected per baryon, which is significantly lower than continuing reactions 
with \hex\ can provide, once (\ref{finallimit}) is implemented.

We shall discuss now some implications of our results. A crucial 
aspect of the cosmological scenario is the initial temperature of the 
Universe $T_{rh}$, which is believed to be generated in the process of 
reheating after inflation. If the $T_{rh}$ is comparable to or larger than 
$\sim 0.05 m_X$, one can count on a sizeable abundance of $\Xmm$, as 
charged particles at these temperatures are guaranteed to stay in 
thermal equilibrium. Subsequent cooling ensures that $\Xmm +X^{++}$ 
annihilation into the SM states (charged particle pairs or photon pairs, for 
example) will lead to a freeze-out abundance of $\Xmm$ that should be 
treated as an {\em initial} abundance for our BBN calculations. Due to 
large mass of $X$ the annihilation cross sections are of a natural electroweak 
size. As a result, despite the annihilation, a significant remainder of 
$X$-particles will be preserved. 

An alternative possibility is the scenario where $T_{rh}$ is below $m_X$, and 
small but non-zero abundance of $X$ occurs due to the ``freeze-in" 
annihilation, SM$\to X^{++}\Xmm$, suppressed by $\exp(-2m_X/T_{rh})$, or 
due to the non-thermal inflaton decay, $I\to {\rm SM} + X^{++}\Xmm$. The 
latter can occur if the mass of the inflaton is larger than $2m_X$. 
Thus, one can identify two broad scenarios for the $\Xmm$ abundance: 
\begin{eqnarray} 
 {A:}~&& {\rm Freeze\mbox{-}out},~X^{++}\Xmm\to{\rm SM};~~ 
\nonumber \\
&& \Xmm/{\rm H} 
= {\cal O}(10^{-2}) \times m_X/{\rm TeV},
\label{foA} \\ 
{B:}~&& {\rm Freeze\mbox{-}in},~{\rm SM}\to X^{++}\Xmm~{\rm or}~I\to {\rm SM} + 
X^{++}\Xmm ;~~ \nonumber \\
&& \Xmm/{\rm H}~\,{\rm model\mbox{-}dependent}\,.
\label{foB}
\end{eqnarray} 
In scenario $B$, the expected abundance of $\Xmm$ is small, but 
is also very model-dependent.

The strength of the BBN limits derived in our work allows us to make 
several important inferences about the cosmological/particle physics 
scenarios related to the mass, lifetime and abundance of $X$ particles.

\begin{itemize}
    
    \item {\em $\Xmm$ within collider reach. } LHC should be able to 
produce $\Xmm+X^{++}$ pairs of TeV rest mass and below. In the 
conventional freeze-out scenario $A$ (\ref{foA}), the BBN limit  
(Fig.~\ref{tau-Y}, lower panel) immediately implies that the lifetime of 
$\Xmm$ should be shorter than about $100$ seconds. 
Such lifetimes would probably be enough for 
subsequent extraction of these particles and study of their decay 
properties, but not enough for their significant accumulations via  
continuous production. Scenario $B$ (\ref{foB}) does not constrain the 
lifetime of $X$ with $m_X\gtrsim 1$ TeV, as $T_{rh}$ can be significantly 
lower \cite{Kudo:2001ie}.

    \item{\em $\Xmm$ with very long lifetimes.} If the lifetime of $\Xmm$ is 
longer than $10^4$\,s, its initial abundance has to be less than 
$10^{-8}$, which is incompatible with the freeze out expectations of 
scenario $A$. In that case, one would have to invoke either some 
additional non-annihilation mechanisms for the $\Xmm$ depletion or 
resort to scenario $B$ with $T_{rh} < 0.05m_X$. If the lifetime of 
$\Xmm$ is comparable to the age of the Universe, {\em and} these particles 
are within 
reach of ${\cal O}(10\,{\rm TeV})$ energy colliders, 
the strong isotopic bound (\ref{He_isotope}) will apply, 
signifying that a ``natural occurrence" of $\Xmm$ on Earth will be very rare.

\item {\em Non-collider probes of stable $\Xmm$.}  
Most of $X^{++}$ or ($^8$Be$\Xmm$) reaching the Earth will chemically 
appear as very heavy helium moving with velocities $v/c \sim 
10^{-3}$. Since the electron binding in helium is rather strong, it is 
expected that the target atoms (in the atmosphere, rock or detector) 
will acquire recoil, and produce ionization. It is debatable whether one 
could re-purpose old searches for magnetic monopoles, see 
Refs.~\cite{Perl:2001xi,Burdin:2014xma}, to look for such signal. At the 
same time, dark matter and coherent neutrino scattering experiments, 
designed to detect keV and sub-keV recoils, will be able to pick the 
collisions of $X^{++}$ or ($^8$Be$\Xmm$) with target atoms. Some of 
these detectors operate near the Earth's surface, so that the velocities 
of $X$-particles will not be moderated by the overburden (see {\em 
e.g.} \cite{CONUS:2020skt}).
Additional opportunities in this direction are provided by large 
scintillator-based neutrino telescopes \cite{Bramante:2018tos}.

\item{\em Anomalous isotopes of elements other than He and H.} In 
addition to $({\rm ^8Be}\Xmm)$ and $({\rm ^6Li}\Xmm)$, which in the case of 
stable or practically stable $\Xmm$ should currently appear as abnormal He and 
H, CBBN predicts that a sizeable fraction of $\Xmm$ will end up bound to 
${\rm ^4He}$. Such $({\rm ^4He}\Xmm)$ states would appear as abnormally heavy 
neutrons. When entering the Earth's atmosphere with CR, they would experience 
fast exothermic charge exchange reactions, transferring their $\Xmm$ to 
nitrogen and oxygen, which would then appear as abnormally heavy  
boron and carbon, respectively. If \hex\ are present at the time of formation 
of the Earth, they may also transfer their $\Xmm$ to heavier elements, 
such as {\em e.g.} iron and silicon, which would then appear as anomalously 
heavy chromium and magnesium, respectively.
For example, as a result of charge exchange one could have the reaction  
$(^4{\rm He}\Xmm)+
{\rm ^{14}N} \to ({\rm ^{14}N}\Xmm)+{\rm ^4He}$ as well as 
$\Xmm$ release via $(^4{\rm He}\Xmm)+{\rm ^{14}N} \to \Xmm+{\rm ^{18}F}$. 
Such reactions should lead to a much higher ionization yield per unit length 
travelled subsequently by $\Xmm$. These new anomalously heavy isotopes can 
also be present in meteorites. Searching for such abnormal isotopes would be 
of considerable interest.

    \item{\em $\Xmm$ as dark matter.} The 
BBN bound (\ref{finallimit}) implies that in order for $\Xmm$ to saturate 
the DM density, its mass has to be close to $10^{10}$\,GeV or above. 
Most of the DM will then be in the form of 
($^8$Be$\Xmm$), \hex\ and $(X^{++}e^-e^-)$. 
With such masses, DM will certainly be out of reach of both current and future 
colliders. While for $m_X\sim 10^4$ GeV or so the bound states 
of $\Xmm$ with Be, Li, N and O can accumulate at the Earth's surface or in the 
upper crust and can be searched for there, in the case of $m_X\gtrsim 10^{10}$ 
GeV they will  likely sink to the Earth's core and therefore will not be 
accessible. In that case, however, their accumulations may still exist 
in meteorites, where also bound states of $\Xmm$ with Fe and Si  
can be present. Thus, should the DM consist of the doubly charged $X$-particles 
and their bound states, the best strategy for foraging for DM would be to  
look for anomalous isotopes of known elements in CR and meteorites.  
 
\end{itemize}

In conclusion, we have derived BBN constraints on abundance {\em vs.} lifetime 
of $\Xmm$ particles through the catalysis of the lithium production. The 
output of lithium in this CBBN scenario can exceed the standard results 
by many orders of magnitude. For parametrically long lifetimes, our 
results imply $\Xmm$/H to be at or below $10^{-9}$ level. If the 
abundance of metastable $\Xmm$ is regulated by the freeze-out processes, 
which can be considered as the most predictive scenario, then the 
lifetimes of $\Xmm$ are limited to about 100 seconds.

\acknowledgments
 We would like to thank Drs. N. Dalal and K. Olive
for useful discussions. M.P. is supported in part by U.S. Department of
Energy Grant No. desc0011842. He also acknowledges the hospitality of
Perimeter Institute in Canada, where part of this work was carried out.
Research at Perimeter Institute is supported by the Government of Canada
through the Department of Innovation, Science and Economic Development
and by the Province of Ontario through the Ministry of Research and
Innovation.

\bibliographystyle{JHEP}
\bibliography{ref1}
\end{document}